\documentclass[twocolumn]{autart}    

\usepackage{amsfonts}
\usepackage{graphics}
\usepackage{epsfig}
\usepackage{mathptmx}
\usepackage{times}
\usepackage{amsmath}
\usepackage{amssymb}
\usepackage{algorithm,algpseudocode}
\usepackage{graphicx}          

\begin{document}

\begin{frontmatter}

\title{Sensor management for multi-target tracking via Multi-Bernoulli filtering\thanksref{footnoteinfo}} 

\thanks[footnoteinfo]{Acknowledgement: The first author is partially supported by Curtin's matching funding to the Australian Research Council Future Fellowship FT0991854 and the second author is supported by Discovery Early Career Researcher Award DE120102388. Corresponding author H.~G.~Hoang.}

\author[a]{Hung Gia Hoang}\ead{hung.hoang@curtin.edu.au},    
\author[a]{Ba Tuong Vo}\ead{ba-tuong.vo@curtin.edu.au},               

\address[a]{Department of Electrical and Computer Engineering, Curtin University, Perth WA 6845, Australia}  

\begin{keyword}                           
multi-target tracking \sep sensor control \sep random finite sets \sep sequential Monte Carlo method               
\end{keyword}                             

\begin{abstract}                          
In multi-object stochastic systems, the issue of sensor management is a theoretically
and computationally challenging problem. In this paper, we present a novel random
finite set (RFS) approach to the multi-target sensor management problem
within the partially observed Markov decision process (POMDP) framework. The
multi-target state is modelled as a multi-Bernoulli RFS, and the multi-Bernoulli filter
is used in conjunction with two different control objectives: maximizing the
expected R\'enyi divergence between the predicted and updated densities, and minimizing
the expected posterior cardinality variance. Numerical studies are presented in two
scenarios where a mobile sensor tracks five moving targets with different levels of
observability.
\end{abstract}

\end{frontmatter}

\section{Introduction}

Multi-target sensor control/management is essentially an optimal non-linear
control problem. The goal of multi-target sensor management is to
\textquotedblleft direct the right sensor on the right platform to the right
target at the right time\textquotedblright\ \cite{Mah03a}. However, the
multi-target sensor control problem differs from the classical control
problem in that it deals with highly complex multi-object stochastic
systems. In multi-object stochastic systems, not only do the number of
objects vary randomly in time, but the measurements are subject to missed
detections and false alarms. This means that the multi-target state and
multi-target observation are inherently finite-set-valued. Consequently,
standard optimal control techniques are not directly applicable \cite{Mah04}%
. Nonetheless, the multi-target sensor scheduling problem can still be cast
in the framework of partially observed Markov decision processes (POMDPs),
where the states and observations are instead finite-set-valued, and control
vectors are drawn from a set of admissible sensor actions based on the
current information states, which are then judged against the values of
an objective function associated with each action \cite{CC08}.

A unified approach to characterizing systems with finite-set-valued states
is the multi-object systems framework, where uncertainty is described by
multi-object probability density functions, and formalized via point process
theory \cite{DJ88,SKM95}, or equivalently by random finite set (RFS) theory
through Mahler's finite set statistics (FISST) \cite{Mah07}. The key
advantage of the RFS based approach is that of a principled framework for
modelling, estimation and control of multi-object systems. In this paper, we
formulate the sensor control problem as a POMDP with an
information-theoretic objective function as well as finite-set-valued states
and measurements. In essence, our approach can be summarized by three basic
steps:

\begin{enumerate}
\item Modelling the sensor and targets as a multi-object stochastic system,
i.e. the multi-target states and multi-target observations as RFSs

\item Propagating the multi-object posterior density recursively in time, or
alternatively a tractable approximation to the posterior;

\item At each time, determining the control action based on optimization of
the reward function over a set of admissible actions.
\end{enumerate}

In the context of single-target tracking, the work in \cite{DVA02} is the
first to propose a practical particle implementation based on the
Kullback-Leibler divergence, and the approach in \cite{SKV07} further
considers the issue of observer trajectory planning. In the more difficult
multi-target context, there are a handful of works falling within the RFS
framework. Using the R\'{e}nyi divergence as the reward function, in \cite%
{RV10} the particle multi-object Bayes filter \cite{VoAES} is used to
propagate the multi-object posterior, while in \cite{RVC11} the particle
probability hypothesis density (PHD) filter \cite{VoAES} is used to
propagate the first moment of the multi-object posterior.

This paper adopts an information theoretic approach for multi-target sensor
control, similar to the approach in \cite{RV10,RVC11}, except that the
Cardinality Balanced Multi-Target Multi-Bernoulli (CB-MeMBer) filter \cite%
{VVC09} is used to propagate a parametrized approximation to the
multi-object posterior. The proposed approach is attractive in that it is
applicable to general non-linear non-Gaussian models, and when coupled with
a particle implementation further reduces the computational load
significantly. Propagating an approximate multi-Bernoulli posterior as
proposed is drastically cheaper than propagating the full multi-object
posterior as in \cite{RV10}, and thus the computation of any associated cost
function using a multi-Bernoulli approximation is generally cheaper than
using the full posterior. While the proposed use of the CB-MeMBer filter
incurs the same complexity as the use of the PHD filter for the same purpose
in \cite{RVC11}, performing state estimation with the former is more
efficient and reliable than the latter because the need for clustering is
eliminated. The work in \cite{RV10,RVC11} also demonstrates that the R\'{e}nyi divergence can be used as a reward function for multi-target sensor
control. In the same regard, the use of the CB-MeMBer filter equally allows
the R\'{e}nyi divergence to be used as a reward function, and further allows a
new type of the reward function to be developed. Since the variance of the
estimated cardinality of a multi-Bernoulli posterior can be evaluated in
closed form \cite{Mah07}, minimizing the cardinality variance can be used as
the control objective, thereby enabling direct control of the cardinality
estimation error.

The main contribution of this paper is a computationally efficient sensor
control algorithm for multiple targets, using the CB-MeMBer filter, as well
as the numerical assessment of two types of control objectives. Our
preliminary result, in particular the idea of using the CB-MeMBer filter,
has been reported in the conference paper \cite{HGH12}. The current paper
provides full details of the algorithm, an alternative cheaper control
objective, and more complete numerical studies.

The organization of the paper is as follows. In Section 2 we review RFS
modelling of multi-object systems and the approximation of the multi-object
posterior density using the CB-MeMBer filter. The two reward functions are
discussed in Section 3 while sequential Monte Carlo (SMC) implementation is
described in Section 4. Section 5 presents simulation results and finally,
Section 6 concludes the paper.


\section{Cardinality Balanced MeMBer filter}

In this section, we summarize the CB-MeMBer filter, the main tool that will
be used throughout the paper. The filter was originally introduced in \cite%
{VVC09} to account for the cardinality bias of the MeMBer filter in \cite%
{Mah07}.

\subsection{General system model}

In contrast with single-object systems where the states and observations are
modelled by random vectors, the states and observations of a multi-object
system are random finite sets of vectors in the single-object state space $%
\mathcal{X}\subseteq \mathbb{R}^{n}$ and single-object observation space
$\mathcal{Z}\subseteq \mathbb{R}^{m}$, respectively:
\begin{align}
\mathbf{X}_{k}& =\{\mathbf{x}_{1}^{k},\ldots ,\mathbf{x}_{n}^{k}\}\in
\mathcal{F}(\mathcal{X}); \\
\mathbf{Z}_{k}& =\{\mathbf{z}_{1}^{k},\ldots ,\mathbf{z}_{m}^{k}\}\in
\mathcal{F}(\mathcal{Z}).
\end{align}%
Here $\mathcal{F}(\mathcal{X})$ and $\mathcal{F}(\mathcal{Z})$ denote the
spaces of all finite subsets of $\mathcal{X}$ and $\mathcal{Z}$. The system is described by the following
probabilistic state space model:
\begin{align}
\mathbf{X}_{k}& \sim \pi _{k|k-1}(\mathbf{X}_{k}|\mathbf{X}_{k-1})
\label{model:trans} \\
\mathbf{Z}_{k}& \sim g_{k}(\mathbf{Z}_{k}|\mathbf{X}_{k})  \label{model:meas}
\end{align}%
where $\mathbf{X}_{k}$ and $\mathbf{Z}_{k}$ respectively are the state and
observation of the system at time $k$. Equation \eqref{model:trans}
describes the system dynamics encapsulating all aspects of object birth,
death and transition while equation \eqref{model:meas} encapsulates all aspects of sensor
detection and false alarms.

Given the system model \eqref{model:trans}-\eqref{model:meas}, the objective
is to determine at each time step $k$ the multi-object posterior probability
density $f_{k}(\mathbf{X}_{k}|\mathbf{Z}_{1:k})$. In the Bayesian filtering
framework, $f_{k}(\mathbf{X}_{k}|\mathbf{Z}_{1:k})$ is obtained through two
steps: time prediction and measurement update \cite{Mah07}. The predicted
density at time $k$ , denoted as $f_{k|k-1}(\mathbf{X}_{k}|\mathbf{Z}%
_{1:k-1})$, is computed by the multi-object Chapman-Kolmogorov equation:
\begin{multline}\label{CKeq}
f_{k|k-1}(\mathbf{X}_k|\mathbf{Z}_{1:k-1}) =\\
 \int \pi_{k|k-1}(\mathbf{X}_k|\mathbf{X}_{k-1})f_{k-1}(\mathbf{X}_{k-1}|\mathbf{Z}_{1:k-1})\delta \mathbf{X}_{k-1}
\end{multline}
where $f_{k-1}(\mathbf{X}_{k-1}|\mathbf{Z}_{1:k-1})$ is the posterior
density from the previous time step $k-1$. When new observations arrive at
the sensor(s), the new posterior density is computed via the multi-object
Bayes rule:
\begin{equation}
f_{k}(\mathbf{X}_{k}|\mathbf{Z}_{1:k})=\frac{g_{k}(\mathbf{Z}_{k}|\mathbf{X}%
_{k})f_{k|k-1}(\mathbf{X}_{k}|\mathbf{Z}_{1:k-1})}{\int g_{k}(\mathbf{Z}_{k}|%
\mathbf{X}_{k})f_{k|k-1}(\mathbf{X}_{k}|\mathbf{Z}_{1:k-1})\delta \mathbf{X}%
_{k}}  \label{Bayesrule}
\end{equation}%
Notice that the integrals in \eqref{CKeq} and \eqref{Bayesrule} are not
ordinary integrals, but are set integrals, and\ that the recursion %
\eqref{CKeq} and \eqref{Bayesrule} has no analytic solution in general. A
sequential Monte Carlo (SMC) implementation of the Bayes multi-object filter
is given in \cite{VoAES}. However, this technique is computationally
prohibitive which at best is able to accommodate a small number of targets.
The multi-target sensor scheduling algorithm proposed in \cite{RV10} employs
this SMC implementation of the multi-object Bayes filter.

Since propagation of the full posterior density given by \eqref{Bayesrule}
is in general intractable, several alternatives have been proposed, which
propagate only summary statistics or important parameters in place of the
full posterior density. For example, the PHD and Cardinalized PHD (CPHD)
filters \cite{Mah03,MahlerCPHD,VoMaGMPHD05,VoAES,VoGaussianCPHD07} propagate
the intensity or first order moment of the posterior density, and were employed by the
multi-target sensor scheduling approach in \cite{RVC11}. An alternative is the
CB-MeMBer filter \cite{Mah07,VVC09}, which propagates a parametrized
multi-Bernoulli approximation of the multi-object posterior density.  The main advantage of the CB-MeMBer approach is its direct
applicability to non-linear non-Gaussian models, which when coupled with an
SMC implementation, obviates the need for the clustering of the particle
population in order to extract estimates.

\subsection{CB-MeMBer Recursion}

We now summarize the recursion for the CB-MeMBer filter. A Bernoulli RFS $\mathbf{X}$ has realizations either as the empty set or a
singleton and is characterized jointly by a probability of existence $r\in
\lbrack 0,1]$ and a probability density $p$. That is, the Bernoulli RFS
takes on a singleton value with probability $r$, and conditional upon
existence, the value of the singleton is distributed according to the
probability density $p$. A multi-Bernoulli RFS is a union of a fixed number $%
M$ of independent Bernoulli RFSs $\mathbf{X}^{(i)}$ with existence
probability $r^{(i)}\in \lbrack 0,1]$ and probability density $p^{(i)}$. A
multi-Bernoulli RFS is completely characterized by the parameter pairs $%
(r^{(i)},p^{(i)})$ and consequently its probability density can be
abbreviated by $\pi _{\mathbf{X}}=\{(r^{(i)},p^{(i)})\}_{i=1}^{M}$. Notice
that the realizations of a multi-Bernoulli RFS are finite sets including the
empty set whose cardinality cannot exceed $M$. As the explicit expression
for the probability density of a multi-Bernoulli RFS is not needed for this
paper, the reader is referred to the original references \cite%
{Mah07,VVC09} for these details.

The premise of the CB-MeMBer filter is that the multi-object posterior
density can be approximated by that of a multi-Bernoulli RFS. Consequently
the CB-MeMBer filter recursively propagates only the parameter set $%
\{(r^{(i)},p^{(i))}\}_{i=1}^{M}$ instead of the full multi-object posterior.

Specifically, if the posterior multi-object density at time $k-1$ can be approximated by a
multi-Bernoulli RFS of the form  $f_{k-1}=\left\{ \left( r_{k-1}^{(i)},p_{k-1}^{(i)}\right) \right\}
_{i=1}^{M_{k-1}}$ then the predicted multi-object density to time $k$ is also that of a
multi-Bernoulli RFS and is given by
\begin{equation}
f_{k|k-1}=\left\{ \left( r_{P,k|k-1}^{(i)},p_{P,k|k-1}^{(i)}\right) \right\}
_{i=1}^{M_{k-1}}\bigcup \left\{ \left( r_{\Gamma ,k}^{(i)},p_{\Gamma
,k}^{(i)}\right) \right\} _{i=1}^{M_{\Gamma ,k}}  \label{pred_den}
\end{equation}%
where $\left\{ \left( r_{\Gamma ,k}^{(i)},p_{\Gamma ,k}^{(i)}\right)
\right\} _{i=1}^{M_{\Gamma ,k}}$ are the parameters of the multi-Bernoulli
RFS of births at time $k$, and
\begin{align}
r_{P,k|k-1}^{(i)}& =r_{k-1}^{(i)}\left\langle
p_{k-1}^{(i)},p_{S,k}\right\rangle  \\
p_{P,k|k-1}^{(i)}(\mathbf{x})& =\frac{\left\langle f_{k|k-1}(\mathbf{x}%
|\cdot ),p_{k-1}^{(i)}(\cdot )p_{S,k}(\cdot )\right\rangle }{\left\langle
p_{k-1}^{(i)},p_{S,k}\right\rangle }
\end{align}%
Here $p_{S,k}(\mathbf{\zeta })$ denotes the probability of object survival
given previous state $\mathbf{\zeta }$; $f_{k|k-1}(\mathbf{%
\cdot }|\mathbf{\zeta })$ denotes the single-object transition density given
previous state $\mathbf{\zeta }$; and $\langle f,g\rangle =\int f(\mathbf{x}%
)g(\mathbf{x})d\mathbf{x}$ is the standard notation for the inner product
between two continuous functions.

For brevity, denote the predicted multi-Bernoulli density specified in %
\eqref{pred_den} by $f_{k|k-1}=\left\{ \left( r_{k|k-1}^{(i)},p_{k|k-1}^{(i)}\right) \right\}
_{i=1}^{M_{k|k-1}}$. Then, the posterior multi-object density at time $k$ can be approximated by
that of a multi-Bernoulli RFS with unbiased cardinality as follows
\begin{equation}
f_{k}=\left\{ \left( r_{L,k}^{(i)},p_{L,k}^{(i)}\right) \right\}
_{i=1}^{M_{k|k-1}}\bigcup \left\{ \left( r_{U,k}(\mathbf{z}),p_{U,k}(\cdot ;\mathbf{z}%
)\right) \right\} _{\mathbf{z}\in \mathbf{Z}_{k}}  \label{upd_den}
\end{equation}%
where
\begin{align}
r_{L,k}^{(i)}& =r_{k|k-1}^{(i)}\frac{1-\langle
p_{k|k-1}^{(i)},p_{D,k}\rangle }{1-r_{k|k-1}^{(i)}\langle
p_{k|k-1}^{(i)},p_{D,k}\rangle }  \label{MeMBer:upd1} \\
p_{L,k}^{(i)}(\mathbf{x})& =p_{k|k-1}^{(i)}(\mathbf{x})\frac{1-p_{D,k}(\mathbf{x})}{%
1-\langle p_{k|k-1}^{(i)},p_{D,k}\rangle } \\
r_{U,k}(\mathbf{z})& =\frac{\sum_{i=1}^{M_{k|k-1}}\frac{%
r_{k|k-1}^{(i)}(1-r_{k|k-1}^{(i)})\langle p_{k|k-1}^{(i)},\psi _{k,\mathbf{z}%
}\rangle }{\left( 1-r_{k|k-1}^{(i)}\langle p_{k|k-1}^{(i)},p_{D,k}\rangle
\right) ^{2}}}{\kappa _{k}(\mathbf{z})+\sum_{i=1}^{M_{k|k-1}}\frac{%
r_{k|k-1}^{(i)}\langle p_{k|k-1}^{(i)},\psi _{k,\mathbf{z}}\rangle }{%
1-r_{k|k-1}^{(i)}\langle p_{k|k-1}^{(i)},p_{D,k}\rangle }} \\
p_{U,k}(\mathbf{x};\mathbf{z})& =\frac{\sum_{i=1}^{M_{k|k-1}}\frac{%
r_{k|k-1}^{(i)}}{1-r_{k|k-1}^{(i)}}p_{k|k-1}^{(i)}(\mathbf{x})\psi _{k,%
\mathbf{z}}(\mathbf{x})}{\sum_{i=1}^{M_{k|k-1}}\frac{r_{k|k-1}^{(i)}}{%
1-r_{k|k-1}^{(i)}}\langle p_{k|k-1}^{(i)},\psi _{k,\mathbf{z}}\rangle } \\
\psi _{k,\mathbf{z}}(\mathbf{x})& =g_{k}(\mathbf{z}|\mathbf{x})p_{D,k}(%
\mathbf{x})  \label{MeMBer:upd2}
\end{align}%
Here $\mathbf{Z}_{k}$ is the measurement set at time $k$, $g_{k}(\cdot |%
\mathbf{x})$ is the single-object likelihood given the current state $%
\mathbf{x}$, $p_{D,k}(\mathbf{x})$ is the probability of object detection at
the current state $\mathbf{x}$, and $\kappa _{k}(\cdot {})$ is the intensity
of Poisson clutter at time $k$.


\section{Objective functions}

The objective function plays a crucial role in sensor control problems as it determines the
manoeuvre of the sensor. Suppose at time $k$ that a control vector $\mathbf{u}_{k}\in \mathbf{U}_{k}$
is applied to the sensor, where $\mathbf{U}_{k}$ is the set of admissible
control vectors. Denote by $\mathcal{R}(\mathbf{u},\mathbf{Z}_{k:k+p})$ the
objective function if we were to apply the control vector $\mathbf{u}$ and
subsequently were to observe the measurement sequence $\mathbf{Z}_{k:k+p}$.
Notice that the reward function depends on $p+1$ future measurements which could be taken to $\infty $. In this paper however, we only consider a single
step ahead (myopic) policy. An established approach to mitigate the presence
of unknown future measurements in the reward function is to take the
expectation of $\mathcal{R}(\mathbf{u},\mathbf{Z}_{k+1})$ over all possible
values of the future measurement $\mathbf{Z}_{k+1}$, i.e. the control vector
$\mathbf{u}_{k}$ is chosen so that $\mathbb{E}\big[\mathcal{R}(\mathbf{u},%
\mathbf{Z}_{k+1})\big]$ is optimal \cite{Mah04,RV10,RVC11}.

In this paper, we present two different types of objective functions. The first approach employs the R\'{e}nyi divergence between the predicted and updated densities, while the second
uses the variance of the maximum a posteriori (MAP) cardinality estimate or
cardinality variance for short.

\subsection{R\'enyi divergence}

The R\'{e}nyi divergence, or alpha divergence, is commonly used as an objective
function in information-theoretic sensor control \cite{RV10,RVC11,HKB08} and encompasses other
information measures such as the Kullback-Leibler divergence or Hellinger
affinity as special cases. Specifically, the R\'{e}nyi divergence between
the predicted and updated densities is defined as follows \cite{HKB08}
\begin{multline}\label{RD}
\mathcal{R}(\mathbf{u},\mathbf{Z}_{1:k+1})=\frac{1}{\alpha -1}\log\int\big[f_{k+1}(\mathbf{X}_{k+1}|\mathbf{Z}_{1:k+1};\mathbf{u})\big]^\alpha\\
[f_{k+1|k}(\mathbf{X}_{k+1}|\mathbf{Z}_{1:k})\big]^{1-\alpha}\delta\mathbf{X}_{k+1}
\end{multline}
where $\alpha \geq 0$ is an adjustable parameter. Let $p(\mathbf{Z}_{k+1}|\mathbf{Z}_k;\mathbf{u})$ designate $\int g_{k+1}(\mathbf{Z}_{k+1}|\mathbf{X}_{k+1};\mathbf{u})f_{k+1|k}(\mathbf{X}_{k+1}|\mathbf{Z}_{1:k})\delta\mathbf{X}_{k+1}$, the R\'{e}nyi divergence can be rewritten as \cite%
{RV10}
\begin{multline}
\mathcal{R}(\mathbf{u},\mathbf{Z}_{k+1})=\frac{1}{\alpha -1}\log \left\{
\frac{1}{\big[p(\mathbf{Z}_{k+1}|\mathbf{Z}_{k};\mathbf{u})\big]^{\alpha }}%
\right.   \label{RD_fin} \\
\left. \int \big[g_{k+1}(\mathbf{Z}_{k+1}|\mathbf{X}_{k+1};\mathbf{u})\big]%
^{\alpha }f_{k+1|k}(\mathbf{X}_{k+1}|\mathbf{Z}_{k})\delta \mathbf{X}%
_{k+1}\right\}
\end{multline}%

In essence, the R\'{e}nyi divergence is a measure of information gain in
terms of dissimilarity between the two densities. An increase in the value
of the R\'{e}nyi divergence can be roughly interpreted as an indication that
more information should be obtained from the future measurements. Thus, the
control vector $\mathbf{u}_{k}$ at time $k$ can be chosen to maximize the
expectation of the R\'{e}nyi divergence over all possible measurements at
time $k+1$: $\displaystyle \mathbf{u}_{k}=\arg \max_{\mathbf{u}\in \mathbf{U}_{k}}\mathbb{E}\big[%
\mathcal{R}(\mathbf{u},\mathbf{Z}_{k+1})\big]$.

The elegance of a R\'{e}nyi divergence based control strategy however
carries several drawbacks. First, it has no analytic solution in general
and thus its computation is potentially expensive. Second, the R\'{e}nyi
divergence is a measure of information gain, but in some abstract
mathematical sense, and it is unclear if or how it translates directly into
practical performance criteria such as cardinality and/or localization
errors.

\subsection{MAP cardinality variance}

In this section, we propose minimizing the estimated cardinality
variance as an alternative objective function, which directly translates to the
minimization of the estimated cardinality error. When the CB-MeMBer filter
is used to propagate the posterior density, the objective function can be
computed analytically, which enables a principled and efficient solution
platform for multi-object sensor control.

Suppose that the posterior density at time $k+1$ is approximated by the
following multi-Bernoulli RFS
\begin{equation}
f_{k+1}(\mathbf{X}_{k+1}|\mathbf{Z}_{k+1};\mathbf{u}_{k})=\left\{
r_{k+1}^{(i)}(\mathbf{u}_{k},\mathbf{Z}_{k+1}),p_{k+1}^{(i)}(\mathbf{u}_{k},%
\mathbf{Z}_{k+1})\right\} _{i=1}^{M}.  \label{mB_RFS}
\end{equation}%
The expected a posteriori (EAP) cardinality estimate and its variance are
\cite{Mah07}:
\begin{align}
\hat{n}_{k+1}^{\text{EAP}}& =\sum_{i=1}^{M}iB_{r_{k+1}^{(1)},\ldots
,r_{k+1}^{(M)}}(i)=\sum_{i=1}^{M}r_{k+1}^{(i)} \label{EAP_card}\\
\big[\sigma _{k+1}^{\text{EAP}}\big]^{2}& =\left[
\sum_{i=1}^{M}i^{2}B_{r_{k+1}^{(1)},\ldots ,r_{k+1}^{(M)}}(i)\right] -\left[
\hat{n}_{k+1}^{\text{EAP}}\right] ^{2}\label{EAP_var}\\
&=\sum_{i=1}^{M}r_{k+1}^{(i)}\left(
1-r_{k+1}^{(i)}\right)
\end{align}%
where $B_{r_{k+1}^{(1)},\ldots ,r_{k+1}^{(M)}}(i)$ denotes the cardinality
distribution of the multi-Bernoulli RFS given in \eqref{mB_RFS}.

The MAP cardinality estimate, $\hat{n}^{\text{MAP}}_{k+1}$ and its variance,
$\big[\sigma^{\text{MAP}}_{k+1}\big]^2$, by definition, are given by
\begin{align}
\hat{n}^{\text{MAP}}_{k+1}&=\arg\max_i B_{r_{k+1}^{(1)},\ldots ,
r_{k+1}^{(M)}}(i)  \label{card_est} \\
\big[\sigma^{\text{MAP}}_{k+1}\big]^2&=\sum_{i=1}^M \left(i-\hat{n}^{\text{%
MAP}}_{k+1}\right)^2B_{r_{k+1}^{(1)},\ldots , r_{k+1}^{(M)}}(i) \\
&=\sum_{i=1}^M i^2B_{r_{k+1}^{(1)},\ldots , r_{k+1}^{(M)}}(i) - %
2\hat{n}^{\text{MAP}}_{k+1}\sum_{i=1}^M iB_{r_{k+1}^{(1)},\ldots ,
r_{k+1}^{(M)}}(i)\notag\\
&\quad + \left[\hat{n}^{\text{MAP}}_{k+1}\right]^2
\label{MAP_var_def}
\end{align}
Substitute \eqref{EAP_card} and \eqref{EAP_var} into \eqref{MAP_var_def}, we obtain
\begin{align}
\big[\sigma _{k+1}^{\text{MAP}}\big]^{2}& =\left( \sigma _{k+1}^{\text{EAP}}\right) ^{2}+\left[ \hat{n}_{k+1}^{\text{%
MAP}}-\hat{n}_{k+1}^{\text{EAP}}\right] ^{2} \\
& =\sum_{i=1}^{M}r_{k+1}^{(i)}\left( 1-r_{k+1}^{(i)}\right) +\left[ \hat{n}%
_{k+1}^{\text{MAP}}-\sum_{i=1}^{M}r_{k+1}^{(i)}\right] ^{2}  \label{sigm_sq}
\end{align}
Minimizing the estimated cardinality variance $\big[\sigma _{k+1}^{\text{MAP}%
}\big]^{2}$ is a natural choice for the control objective, as it drives the
sensor towards positions that yield better MAP estimates of the number of
targets. Note however, that even though the objective function $\mathcal{R}(%
\mathbf{u}_{k},\mathbf{Z}_{k+1})=\big[\hat{\sigma}_{k+1}^{\text{MAP}}(%
\mathbf{u}_{k},\mathbf{Z}_{k+1})\big]^{2}$ can be computed analytically, it
still depends on future measurements. Consequently an approximation is
needed to determine the optimal control since it involves the expectation $%
\mathbb{E}\left[ \mathcal{R}(\mathbf{u}_{k},\mathbf{Z}_{k+1})\right] $ over
all possible values of future measurements $\mathbf{Z}_{k+1}$:
\begin{equation}
\mathbf{u}_{k}=\arg \min_{\mathbf{u}\in \mathbf{U}_{k}}\mathbb{E}\left\{ %
\left[ \hat{\sigma}_{k+1}^{\text{MAP}}(\mathbf{u}_{k},\mathbf{Z}_{k+1})%
\right] ^{2}\right\}   \label{obj2}
\end{equation}
Although the use of the cardinality variance offers several advantages over
the R\'{e}nyi divergence as a reward function, it also has drawbacks, most
notably that the localization error is nowhere accounted for in the reward
function. As a result, this control strategy is expected to perform well if
the sensor gives reasonable information on the actual state variables of the
targets. However, if the sensor does not provide sufficient information on
the state variables, for example bearing or range only sensors, the
performance degrades in pathological cases as we will illustrate with an
example later.

\section{SMC implementation}

Since there are no general analytic solutions to the CB-MeMBer recursions %
\eqref{pred_den}-\eqref{upd_den}, or the R\'{e}nyi divergence equation %
\eqref{RD}, it is necessary to employ numerical approximations to implement
both the filter and control algorithms. This paper adapts the SMC method
given in \cite{RAG04} in order to perform multi-object sensor control with
the CB-MeMBer filter. We will show that employing the MAP cardinality variance as
the objective function enables a fast SMC implementation which does not require
sampling on the multi-target state space.

The SMC implementations for each control strategy are similar and differ only in the subroutines used to compute the optimal control command. Algorithm~\ref{alg1} gives the pseudo code for a general sensor management algorithm. Details and pseudo codes for computation of the optimal control command indicated in Step 5 are given in the following subsections based on the  R\'enyi divergence and cardinality variance objective functions.
\renewcommand{\algorithmicrequire}{\textbf{Inputs:}} \renewcommand{%
\algorithmicensure}{\textbf{Outputs:}}
\begin{algorithm}[htb]
\caption{General multi-Bernoulli sensor management}
\label{alg1}
\begin{algorithmic}[1]
    \State k=0, initialising the multi-Bernoulli density $\left\lbrace r_0^{(i)}, \left\lbrace w_0^{(i,j)},\mathbf{x}_0^{(i,j)}\right\rbrace_{j=1}^{L_0^{(i)}}\right\rbrace_{i=1}^{M_0}$ \Comment initialisation
    \Repeat
    \State k=k+1 \Comment time step increment
    \State Calculate the predicted density 
     \Statex \quad \;$\left\lbrace r_{k|k-1}^{(i)}, \left\lbrace w_{k|k-1}^{(i,j)},\mathbf{x}_{k|k-1}^{(i,j)}\right\rbrace_{j=1}^{L_{k|k-1}^{(i)}}\right\rbrace_{i=1}^{M_{k|k-1}}$ 
    \State Compute the optimal control command $\mathbf{u}_k^\star$
    \State Apply $\mathbf{u}_k^\star$ and obtain the measurement set $\mathbf{Z}_{k}$
    \State Compute the data update posterior
    \Statex \quad \;$\left\lbrace r_{k}^{(i)}, \left\lbrace w_{k}^{(i,j)},\mathbf{x}_{k}^{(i,j)}\right\rbrace_{j=1}^{L_{k}^{(i)}}\right\rbrace_{i=1}^{M_{k}}$
    \Until{stop}
\end{algorithmic}
\end{algorithm}

\subsection{SMC implementation for maximizing R\'enyi divergence}
Let the predicted multi-object density at time $k+1$ be represented by the following SMC form
\begin{equation}
f_{k+1|k}(\mathbf{X})\simeq\sum_{i=1}^{S}w^{i}\delta _{\mathbf{X}_{k+1}^{i}}(%
\mathbf{X})  \label{pred_SMC}
\end{equation}%
where each multi-object particle $\mathbf{X}_{k+1}^{i}$ is sampled from a
proposal (importance) density $q_{k+1|k}(\cdot |\mathbf{X}_{k}^{i},\mathbf{Z}%
_{k})$, and $w^{i}$ is the weight associated with it. The R\'enyi divergence, $\mathcal{R}(\mathbf{u},\mathbf{Z}_{k+1})$, is obtained by substituting \eqref{pred_SMC} for \eqref{RD_fin} 
\begin{equation}
\mathcal{R}(\mathbf{u},\mathbf{Z}_{k+1})=\frac{1}{\alpha -1}\log \frac{%
\sum_{i=1}^{S}w^{i}\big[g_{k+1}(\mathbf{Z}_{k+1}|\mathbf{X}_{k+1}^{i};\mathbf{u})\big]^{\alpha }}{\left[ \sum_{i=1}^{S}w^{i}g_{k+1}(\mathbf{Z}%
_{k+1}|\mathbf{X}_{k+1}^{i};\mathbf{u})\right] ^{\alpha }}  \label{RD_SMC}
\end{equation}

There are two approaches to computing the expectation of \eqref{RD_SMC}. The
first is to generate an ensemble of measurement sets $\mathbf{Z}_{k+1}$ for
a given clutter intensity and detection rate via the measurement model \cite%
{RV10}. In this paper, we adopt a computationally cheaper approach, which
uses the ideal predicted measurement set as in \cite{Mah04,RVC11}. Instead
of using a set of samples $\left\{ \mathbf{Z}_{k+1}^{(\zeta)}\right\}
_{\zeta=1}^{T}$ from the measurement space, we generate only one future
measurement, assuming zero clutter and unity detection rate, and based on
the estimated states from the predicted density. Although this is similar to
the approach in \cite{RVC11} which uses the PHD filter to propagate the
posterior density, the proposed use of the particle CB-MeMBer filter to
propagate the posterior density generally results in more efficient and
reliable state estimates, compared to that of PHD filter in \cite{RVC11}. The pseudo code for the control subroutine based on an SMC\
implementation of the CB-MeMBer filter and a R\'{e}nyi divergence based objective function is
provided in Algorithm~\ref{alg2}.
\begin{algorithm}[htb]
\caption{R\'enyi divergence based sensor control subroutine}
\label{alg2}
\begin{algorithmic}[1]
    \Require Predicted pdf $\left\lbrace r_{k|k-1}^{(i)}, \left\lbrace w_{k|k-1}^{(i,j)},\mathbf{x}_{k|k-1}^{(i,j)}\right\rbrace_{j=1}^{L_{k|k-1}^{(i)}}\right\rbrace_{i=1}^{M_{k|k-1}}$ and $n$ - number of Monte Carlo samples of the states 
    \Ensure $\mathbf{u}_k^\star$
    \State Sampling the multi-object state space
    \Statex \quad \; $\left\{\mathbf{\bar{X}}_{k}^{(\iota)}\right\}_{\iota=1}^{S} \sim \left\lbrace r_{k|k-1}^{(i)}, \left\lbrace w_{k|k-1}^{(i,j)},\mathbf{x}_{k|k-1}^{(i,j)}\right\rbrace_{j=1}^{L_{k|k-1}^{(i)}}\right\rbrace_{i=1}^{M_{k|k-1}}$
    \State Calculate $\hat{n}^{MAP}_{k|k-1}$ and $\mathbf{\hat{X}}_{k|k-1}=\left\{\mathbf{\hat{x}}_{k|k-1}^{(i)}\right\}_{i=1}^{\hat{n}^{MAP}_{k|k-1}}$
    \State Calculate admissible control commands
    $\mathbf{U}_k=\left\{\mathbf{u}_k^{(\ell)}\right\}_{\ell=1}^{N_U}$
    \For{$\ell \gets 1, N_U$}
    \State Get $\mathbf{\hat{Z}}_{k}^{(\ell)}$ based on $\mathbf{u}_k^{(\ell)}$ and $\mathbf{\hat{X}}_{k|k-1}$, with $p_D=1$
    \Statex \quad\; and $\lambda=0$
    \State Calculate hypothesized multi-object likelihoods
    \Statex \quad\;$\left\lbrace g\left(\mathbf{\hat{Z}}_{k}^{(\ell)}|\mathbf{\bar{X}}_{k}^{(\iota)}\right)\right\rbrace_{\iota=1}^{S}$
    \State Compute $\mathcal{R}\left(\mathbf{u}_k^{(\ell)}\right)$ based on $\left\lbrace g\left(\mathbf{\hat{Z}}_{k}^{(\ell)}|\mathbf{\bar{X}}_{k}^{(\iota)}\right)\right\rbrace_{\iota=1}^{S}$
    \EndFor
    \State $\mathbf{u}_k^\star \gets \arg\max \mathcal{R}\left(\mathbf{u}_k^{(\ell)}\right)$
\end{algorithmic}
\end{algorithm}

\subsection{SMC implementation for minimizing cardinality variance}

Although the cardinality variance can be computed analytically with a CB-MeMBer filter, it is still
necessary to employ Monte Carlo approximation to side step the
presence of future measurements in the expectation formulation \eqref{obj2}. Let $\left\{ \bar{%
\mathbf{Z}}_{k+1}^{(\zeta)}\right\} _{j=1}^{T}$ be random samples of the
multi-object measurement space generated at time $k$, then the expectation
of the MAP cardinality variance can be approximated by
\begin{equation}
\mathbb{E}\left\{ \left[ \sigma _{k+1}^{MAP}\left( \mathbf{u}_{k},\mathbf{Z}%
_{k+1}\right) \right] ^{2}\right\} \simeq \frac{1}{T}\sum_{\zeta=1}^{T}%
\sigma _{k+1}^{MAP}\left( \mathbf{u}_{k},\bar{\mathbf{Z}}_{k+1}^{(\zeta)}\right)
\label{SMC_sigm}
\end{equation}%
where $\left[ \sigma _{k+1}^{MAP}\left( \mathbf{u}_{k},\bar{\mathbf{Z}}%
_{k+1}^{(\zeta)}\right) \right] ^{2}$ is an instance of the cardinality variance at time $k+1$ based on the hypothesized future measurement $\bar{\mathbf{Z}}_{k+1}^{(\zeta)}$. Thus, the approximation \eqref{SMC_sigm} converges to the true value of $\mathbb{E}\left\{ \left[ \sigma _{k+1}^{MAP}\left( \mathbf{u}_{k},\mathbf{Z}_{k+1}\right) \right] ^{2}\right\}$ as $T \rightarrow \infty $. The pseudo code for this
control strategy is shown in Algorithm~\ref{alg3}.
\begin{algorithm}[htb]
\caption{Cardinality variance based sensor control subroutine (using measurement sampling)}
\label{alg3}
\begin{algorithmic}[1]
    \Require Predicted pdf $\left\lbrace r_{k|k-1}^{(i)}, \left\lbrace w_{k|k-1}^{(i,j)},\mathbf{x}_{k|k-1}^{(i,j)}\right\rbrace_{j=1}^{L_{k|k-1}^{(i)}}\right\rbrace_{i=1}^{M_{k|k-1}}$ and $m$ - number of MC samples of future measurements
    \Ensure $\mathbf{u}_k^\star$
    \State Sampling the multi-object state space
    \Statex \quad \; $\left\{\mathbf{\bar{X}}_{k}^{(\iota)}\right\}_{\iota=1}^{S} \sim \left\lbrace r_{k|k-1}^{(i)}, \left\lbrace w_{k|k-1}^{(i,j)},\mathbf{x}_{k|k-1}^{(i,j)}\right\rbrace_{j=1}^{L_{k|k-1}^{(i)}}\right\rbrace_{i=1}^{M_{k|k-1}}$
    \State Calculate admissible control commands $\mathbf{U}_k=\left\{\mathbf{u}_k^{(\ell)}\right\}_{\ell=1}^{N_U}$
    \For{$\ell \gets 1, N_U$}
    \State Get $\left\{\mathbf{\bar{Z}}_{k}^{(\ell,\zeta)}\right\}_{\zeta=1}^{T}$ based on $\left\{\mathbf{\bar{X}}_{k}^{(\iota)}\right\}_{\iota=1}^{S}$, $p_D(\cdot)$, $\kappa(\cdot)$
    \For{$\zeta \gets 1, T$}
    \State Compute $\left[\sigma^{MAP,(\zeta)}\left(\mathbf{u}_k^{(\ell)}\right)\right]^2$ using $\mathbf{\bar{Z}}_{k}^{(\ell,\zeta)}$ and
    \Statex \qquad\qquad\; $\left\{\mathbf{\bar{X}}_{k}^{(\iota)}\right\}_{\iota=1}^{S}$
    \EndFor
    \State $\mathbb{E}\left[\mathcal{R}\left(\mathbf{u}_k^{(\ell)}\right)\right] = \frac{1}{T}\sum_{\zeta=1}^{T} \left[\sigma^{MAP,(\zeta)}\left(\mathbf{u}_k^{(\ell)}\right)\right]^2$
     \EndFor
    \State $\mathbf{u}_k^\star \gets \arg\min \mathbb{E}\left[\mathcal{R}\left(\mathbf{u}_k^{(\ell)}\right)\right]$
\end{algorithmic}
\end{algorithm}

While the algorithm presented in Algorithm~\ref{alg3} is conceptually straightforward to implement, it is still computationally expensive, as it depends on two long nested for loops to produce samples of the cardinality variance for each control command. Nonetheless, by exploiting the
efficiency and reliability of the CB-MeMBer filter's state estimation, we can develop a
simple method to drastically reduce the computational workload.
In essence, instead of using a large amount of particles from the sampled multi-object states $\left\{\mathbf{\bar{X}}_{k}^{(\iota)}\right\}_{\iota=1}^{S}$ to represent the predicted density, we truncate the predicted density to $\hat{n}_{k|k-1}^{MAP}$ Bernoulli components with highest existence probabilities, and
use the estimated state $\mathbf{\hat{X}}_{k|k-1}=\left\{ \mathbf{\hat{x}}%
_{k|k-1}^{(i)}\right\} _{i=1}^{\hat{n}_{k|k-1}^{MAP}}$ as the spatial
distributions, i.e. the predicted density is now represented by $\left\{
\hat{r}_{k|k-1}^{(i)},\left\{ 1,{\mathbf{\hat{x}}_{k|k-1}^{(i)}}\right\}
\right\} _{i=1}^{\hat{n}_{k|k-1}^{MAP}}$. The estimated state $\mathbf{\hat{X}}_{k|k-1}$ is also used to generate one ideal future measurement set in order to compute the expectation of the cardinality variance. The advantages of the proposed approach are two-fold:
\begin{itemize}
\item As compared to the CB-MeMBer based strategies introduced in Algorithm~\ref{alg2} and Algorithm~\ref{alg3}, computational workload is significantly reduced as the number of particles is minimal;
\item As compared to the PHD based counterpart \cite{RVC11}, tracking performance is improved because the need for clustering techniques is eliminated.
\end{itemize}
The pseudo code for this method is presented in Algorithm~\ref{alg4}.
\begin{algorithm}[htb]
\caption{Cardinality variance based sensor control subroutine (non-sampling method)}
\label{alg4}
\begin{algorithmic}[1]
    \Require Predicted pdf $\left\lbrace r_{k|k-1}^{(i)}, \left\lbrace w_{k|k-1}^{(i,j)},\mathbf{x}_{k|k-1}^{(i,j)}\right\rbrace_{j=1}^{L_{k|k-1}^{(i)}}\right\rbrace_{i=1}^{M_{k|k-1}}$
    \Ensure $\mathbf{u}_k^\star$
    \State Calculate $\hat{n}^{MAP}_{k|k-1}$ and $\mathbf{\hat{X}}_{k|k-1}=\left\{\mathbf{\hat{x}}_{k|k-1}^{(i)}\right\}_{i=1}^{\hat{n}^{MAP}_{k|k-1}}$
    \State Calculate admissible control commands $\mathbf{U}_k=\left\{\mathbf{u}_k^{(\ell)}\right\}_{\ell=1}^{N_U}$
    \For{$\ell \gets 1, N_U$}
    \State Get $\mathbf{\hat{Z}}_{k}^{(\ell)}$ based on $\mathbf{u}_k^{(\ell)}$ and $\mathbf{\hat{X}}_{k|k-1}$, with $p_D=1$
    \Statex \quad\; and $\lambda=0$
    \State Compute $\left[\sigma^{MAP}\left(\mathbf{u}_k^{(\ell)}\right)\right]^2$  via $\mathbf{\hat{Z}}_{k}^{(j)}$ and 
    \Statex \quad\; $\left\{\mathbf{\hat{x}}_{k|k-1}^{(i)}\right\}_{i=1}^{\hat{n}^{MAP}_{k|k-1}}$
    \EndFor
    \State $\mathbf{u}_k^\star \gets \arg\min \left[\sigma^{MAP}\left(\mathbf{u}_k^{(\ell)}\right)\right]^2$
\end{algorithmic}
\end{algorithm}


\section{Simulation results}

In order to demonstrate the proposed approach, we use a numerical example
adapted from \cite{RVC11}, where a mobile range and bearing sensor is
tracking 5 moving targets. The surveillance area is a square of dimensions $%
1000m\times 1000m$. Each target in this area is characterized by a
single-object state of the form $\mathbf{x}=[x\;\,y\;\,\dot{x}\;\,\dot{y}]^{T}$, where $[x\;\,y]^{T}$ is the position and $%
[\dot{x}\;\,\dot{y}]^{T}$ is the velocity of the target.

The set of admissible control vectors $\mathbb{U}_{k}$ is computed as follows. If
the current position of the sensor is $\mathbf{s}_{k}=[s^x_k\;\,s^y_k]^{T}$, the set of all
possible one-step ahead control actions is:
\begin{equation*}
\mathbf{U}_{k}=\left\{ (s^x_{k}+j\Delta _{R}\cos (\ell \Delta _{\theta
}),s^y_{k}+j\Delta _{R}\sin (\ell \Delta _{\theta }))\right\} _{j=0,\ldots
,N_{R}}^{\ell =0,\ldots ,N_{\theta }}
\end{equation*}%
where $\Delta _{\theta }=2\pi /N_{\theta }$ and $\Delta _{R}=50m$ is the
radial step size. Other parameters are $N_{R}=2$, $N_{\theta }=8$, and $%
\Delta _{R}=50m$. The observer is
always kept inside the surveillance area by setting the reward function
associated with control vectors outside the area to $-\infty $.

If the sensor is at position $\mathbf{s}$, it detects an object at position $\mathbf{p}=\mathbf{H}\mathbf{x}$ with probability
\begin{equation}
p_{D}(\mathbf{x},\mathbf{s})=%
\begin{cases}
0.99, & \Vert \mathbf{p}-\mathbf{s}\Vert \leq R_{0} \\
\max \{0,0.99-(\Vert \mathbf{p}-\mathbf{s}\Vert -R_{0})\hbar \}, & \Vert
\mathbf{p}-\mathbf{s}\Vert >R_{0}%
\end{cases}%
\end{equation}%
where $\mathbf{H}=\begin{bmatrix}1 &0 &0 &0\\0 &1 &0 &0\end{bmatrix}$, $\Vert \mathbf{p}-\mathbf{s}\Vert $ is the Euclidean distance between the sensor and object, and $R_{0}=300m$, $\hbar =0.0005m^{-1}$.

The single-object transition density is $\pi _{k|k-1}(\mathbf{x}_{k}|\mathbf{%
x}_{k-1})=\mathcal{N}(\mathbf{x}_{k};\mathbf{F}\mathbf{x}_{k-1},\mathbf{Q})$%
, where
\begin{equation*}
\mathbf{F}=%
\begin{bmatrix}
1 & 0 & T & 0 \\
0 & 1 & 0 & T \\
0 & 0 & 1 & 0 \\
0 & 0 & 0 & 1%
\end{bmatrix}%
,\;\mathbf{Q}=27%
\begin{bmatrix}
T^{3} & 0 & \frac{T^{2}}{54} & 0 \\
0 & T^{3} & 0 & \frac{T^{2}}{54} \\
\frac{T^{2}}{54} & 0 & \frac{T}{81} & 0 \\
0 & \frac{T^{2}}{54} & 0 & \frac{T}{81}%
\end{bmatrix}%
\end{equation*}%
with $T=1s$. Measurements are noisy bearing and range returns according to
the single-object likelihood $g(\mathbf{z}|\mathbf{x};\mathbf{s})=\mathcal{N}(\mathbf{z}%
;[\Vert \mathbf{p}-\mathbf{s}\Vert ;0],\Sigma ^{T}\Sigma )$, where $\Sigma =%
\begin{bmatrix}
\sigma _{\zeta } & 0 \\
0 & \sigma _{\phi }%
\end{bmatrix}%
$ with
\begin{align}
\sigma _{\zeta }& =\sigma _{0}+\beta _{\zeta }\Vert \mathbf{p}-\mathbf{s}%
\Vert ^{2}, \\
\sigma _{\phi }& =\phi _{0}+\beta _{\phi }\Vert \mathbf{p}-\mathbf{s}\Vert ,
\end{align}%
and $\sigma _{0}=1m$, $\beta _{\zeta }=5\cdot 10^{-5}m^{-1}$, $\phi _{0}=%
\frac{\pi }{180}rad$, $\beta _{\phi }=10^{-5}rad{\cdot }m^{-1}$. Clutter is
modelled by a Poisson RFS whose intensity is $\kappa (\mathbf{z})=\lambda \cdot
c(\mathbf{z})$ with $\lambda =5$ and $c(\mathbf{z})=\mathcal{U(}\mathbf{z}%
;[0,R_{\text{max}}]\times \lbrack 0,\frac{\pi }{2}])$, where $R_{\text{max}}$
is the maximum distance from the sensor location to the vertices of the
surveillance area. One target dies at time $k=19$ and a new target is born away from the existing cluster of targets at time $k=27$.

\begin{figure}[htb]
\centering
\includegraphics[scale=.52]{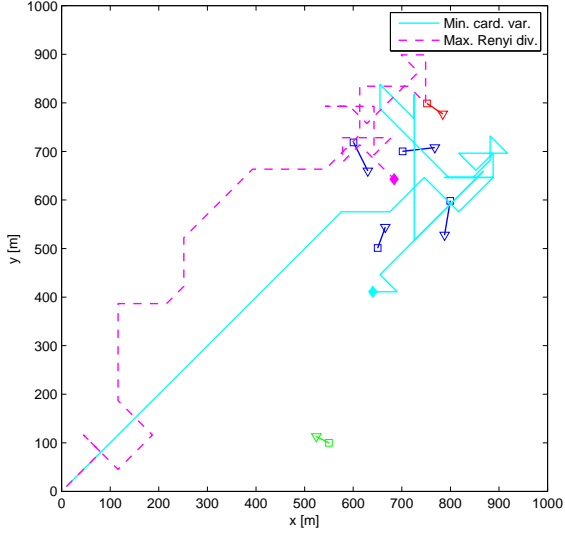}
\caption{Typical sensor manoeuvres. Target start and stop positions are marked by $\square$ and $\nabla$, respectively. The red target died at $k=19$ while the green target was born at $k=27$.}
\label{fig1}
\end{figure}

For this scenario, and for the purposes of performing sensor control with either the R%
\'{e}nyi divergence or cardinality variance based cost functions, the range
and bearing information acquired by the sensor is considered to be generally
informative. Consequently, it is expected that a \textquotedblleft\
good\textquotedblright\ control policy should, intuitively speaking, move
the sensor towards the targets, and remain in their vicinity in order to
obtain high detection probabilities and low measurement noise. Fig.~\ref{fig1} illustrates the
typical resultant sensor trajectories for the two proposed control
strategies. It can be seen that both control policies result in sensor behaviours that would be
expected of a ``good'' control policy as hypothesized above. Minimizing the cardinality variance appears to result in
smoother changes in the sensor trajectory while maximizing R\'{e}nyi divergence
appears to permit impulsive jumps in the sensor position.

However, the two control strategies appear to respond differently to target births and deaths. The R\'{e}nyi divergence based control prefers for the sensor to stay close to confirmed targets, reacting immediately to a target death nearby, but only changing positions slightly when there is a target birth far away. This is because the R\'{e}nyi divergence cost seeks the ``best'' overall information taken across all targets, which is small if the sensor were to move towards a single unconfirmed target located at a distance from the sensor, but much larger if the sensor were to stay in the vicinity of multiple targets which are already confirmed. Moreover moving away from multiple confirmed targets and towards a possible unconfirmed target may even result in a decrease in the overall information. Hence the sensor remains in the vicinity of existing targets. The cardinality variance based strategy on the other hand moves the sensor to middle of the estimated target positions regardless of births of deaths. This is because the cardinality variance based control does not take into target localization error, and thus seeks to optimize the cardinality estimate even if position estimates are poor. Consequently the sensor is driven to maximize the detection probability and minimize the measurement noise across all targets which is generally achieved by positioning itself in the middle of suspected detections.

We proceed to illustrate and compare the performance of the R\'{e}nyi
divergence and cardinality variance based sensor control strategies. In the case of the R\'{e}nyi
divergence, the implementation shown in Algorithm~\ref{alg2} is used to demonstrate the proposed CB-MeMBer filter based approach. For comparison, two PHD filter based algorithms are employed. The first approach uses a standard SMC implementation, which performs state estimation through k-means clustering. The second approach, as proposed in \cite{RVC11} sidesteps clustering with measurement driven state estimation techniques. In the case of the cardinality variance based control strategies, the two implementations shown in Algorithm~\ref{alg3} and~\ref{alg4} are trialled and compared. All algorithms were implemented in MATLAB R2010b on a laptop with an Intel Core i5-3360 CPU and 8GB of RAM. The average run time for the R\'{e}nyi divergence based strategies are 0.99 seconds (CB-MeMBer), 2.46 seconds (PHD without clustering), and  2.51 seconds (PHD with $k$-means clustering) while those for cardinality variance based strategies are 0.61 seconds (non sampling) and 41.35 seconds (with sampling). As expected, the non sampling CB-MeMBer based strategy is the fastest.

\begin{figure}[htb]
\centering
\includegraphics[scale=.49]{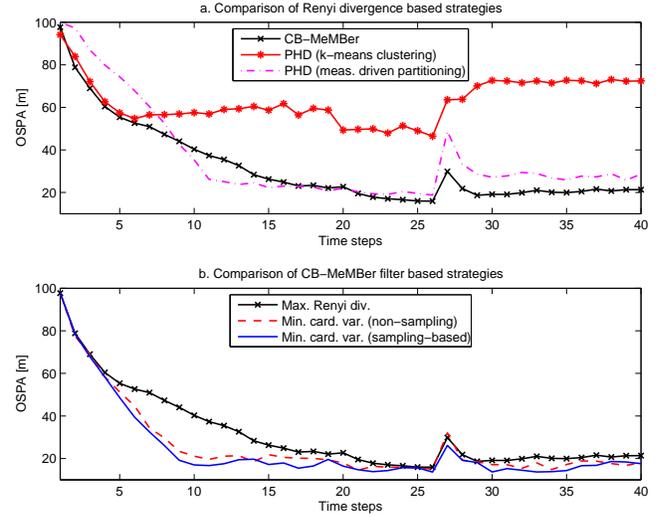}
\caption{Comparison of OSPA by different methods (range bearing)}
\label{fig2}
\end{figure}

Fig.~\ref{fig2} shows the averaged Optimal SubPattern Assignment (OSPA) metric \cite{SVV08} (with parameters $p=2$, $c=100m$) over 200 Monte Carlo runs for different control strategies. The OSPA curves in Fig.~\ref{fig2}a suggest that the performance of PHD based strategies relies heavily on the partitioning of the particle population. While k-means clustering is highly unreliable, the measurement driven estimation \cite{RVC11} is more consistent. Furthermore, using the same R\'enyi divergence objective function, the CB-MeMBer filter based strategy outperforms the PHD filter based counterparts in terms of miss distance by approximate 10m in steady state. On the other hand, among CB-MeMBer based strategies, the OSPA curves of the cardinality variance based approaches decrease faster than that of the R\'{e}nyi divergence based counterpart and are marginally lower in steady state as shown in Fig.~\ref{fig2}b. This is because the cardinality variance based control strategies tend to drive the sensor more quickly towards targets which are initially present and tend to respond more sensitively to the birth of new targets. In contrast the R\'enyi divergence based control strategies tend to favour existing targets as they are on average more informative in terms of the objective function. As expected however, among the cardinality variance based approaches, the computational savings realized by the non sampling approach incur a slightly higher estimation error. To sum up, the results of Figure~\ref{fig2} indicate that, if the sensor obtains sufficient information on the target states, all CB-MeMBer based control strategies are effective and outperform the PHD based counterparts.

We now present a pathological example where the performance of the sensor
control algorithm deteriorates significantly. Instead of bearings and range
measurements, we now consider range only measurements and assume no target birth or death. It is expected that
both the R\'{e}nyi divergence and cardinality variance based strategies
would both perform worse, since the range only measurements are much less
informative, and actually result in a lower target \textquotedblleft
observability\textquotedblright . The single-object measurement model is now
given by $g(\mathbf{z}|\mathbf{x};\mathbf{s})=\mathcal{N}(\mathbf{z};\Vert \mathbf{p}-%
\mathbf{s}\Vert ,\sigma _{\zeta }^{2})$. Clutter is modelled by a Poisson
RFS with intensity $\kappa (\mathbf{z})=\lambda \cdot c(\mathbf{z})$ with $c(%
\mathbf{z})=\mathcal{U(}\mathbf{z};[0,R_{\text{max}}])$. All other
parameters are the same as those in the previous scenario.

\begin{figure}[htb]
\centering
\includegraphics[scale=.48]{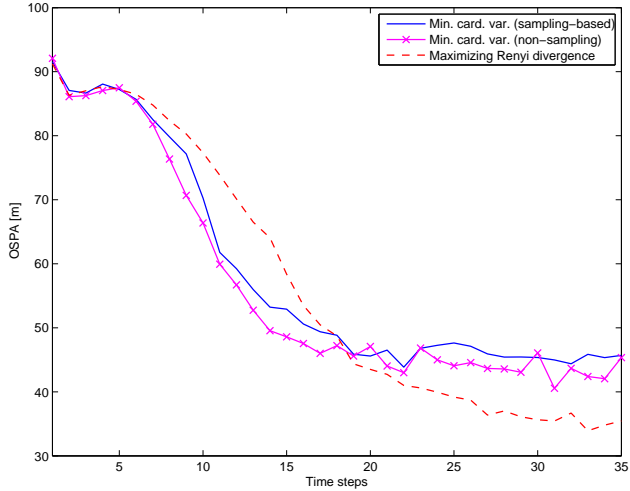}
\caption{Comparison of OSPA by different methods (range
only) }
\label{fig4}
\end{figure}

Fig.~\ref{fig4} shows the averaged OSPA distance (with parameters $p=2$, $%
c=100m$) corresponding to the two objective functions after 200 Monte Carlo
runs. It is clear that the R\'{e}nyi divergence based strategy performs
better, although neither control strategy achieves a sufficiently low OSPA
error. The under-performance of both strategies can be explained by the lack
of observability of the full states, as we now have range-only measurements.
Minimizing the cardinality variance tends to drive the observer to a
position where all the targets have roughly the same range so that the
sensor can detect each of the targets equally well. However for range only
sensors, this results in difficulty resolving the targets. On the other
hand, an information theoretic criterion such as the R\'{e}nyi divergence tends to
account for both localization and cardinality criteria (albeit indirectly),
and at least in this particular scenario, results in a different control
policy with a lower error. Thus the \textquotedblleft
observability\textquotedblright\ of the targets is clearly a factor in
determining the estimation performance of a particular control policy in
different scenarios. Further work is required to investigate the notion of
\textquotedblleft observability\textquotedblright\ in a multi-target
situation, and to develop a deeper understanding of what types of control
policies are appropriate for different scenarios.


\section{Conclusions and future works}

The paper has presented a computationally tractable solution to the
multi-target sensor management problem by combining POMDP theory with RFS or
FISST based modelling. The CB-MeMBer filter was used to efficiently but
approximately propagate a parametrized representation of the multi-object
posterior density, which was then used to calculate a reward function in
order to determine the action of the sensor. Using the particle CB-MeMBer
filter not only allowed a conventional reward function, such as the R\'{e}%
nyi divergence, to be calculated efficiently, but also enabled the estimated
cardinality variance, which admits an analytic solution and provides direct
control of cardinality estimation error, to be used as an objective function. In
the latter case, a computationally fast control algorithm that does not
require state space sampling has further been proposed. Numerical examples
were presented, showing that when the sensor measurements are reasonably
informative, with a high \textquotedblleft observability\textquotedblright\
of the targets, both proposed control strategies work well. However, in
pathological cases where there is low \textquotedblleft
observability\textquotedblright\ of targets, even though performance of both
control policies degrades, the decrease in performance is noticeably worse
for the cardinality variance based approach compared to the R\'{e}nyi
divergence based approach. Thus the \textquotedblleft
observability\textquotedblright\ of the targets is clearly a factor in
determining the performance of a particular control policy in different
scenarios. Further work is required to investigate the notion of
\textquotedblleft observability\textquotedblright\ in a multi-target
situation, and to develop a deeper understanding of what types of control
policies are appropriate for different scenarios.


\begin{thebibliography}{99}
\bibitem{CC08}
D.~A. Castan\'on and L.~Carin.
\newblock Stochastic control theory for sensor management.
\newblock In A.~O. Hero, D.~A. Castan\'on, D.~Cochran, and K.~Kastella,
  editors, {\em Foundations and Applications of Sensor Management}, chapter~2,
  pages 7--32. Springer, 2008.

\bibitem{DJ88}
D.~Daley and D.~Vere-Jones.
\newblock {\em An introduction to the theory of point processes}.
\newblock Springer-Verlag, 1988.

\bibitem{DVA02}
A.~Doucet, B.-N. Vo, C.~Andrieu, and M.~Davy.
\newblock Particle filtering for multi-target tracking and sensor management.
\newblock In {\em Proc. 5th Annual Conference on Information Fusion (FUSION
  2002)}, volume~1, pages 474--481, Annapolis Maryland, 2002.

\bibitem{HKB08}
A.~O. Hero, C.~M. Kreucher, and D.~Blatt.
\newblock Information theoretic approaches to sensor management.
\newblock In A.~O. Hero, D.~A. Castan\'on, D.~Cochran, and K.~Kastella,
  editors, {\em Foundations and applications of sensor management}, chapter~3,
  pages 33--57. Springer, 2008.

\bibitem{HGH12}
H.~G. Hoang.
\newblock Control of a mobile sensor for multi-target tracking using
  {Multi-Target/Object Multi-Bernoulli} filter.
\newblock In {\em Proc. International Conference on Control, Automation and
  Information Sciences (ICCAIS 2012)}, pages 7--12, Ho Chi Minh City, Vietnam,
  2012.

\bibitem{Mah03}
R.~Mahler.
\newblock Multi-target {Bayes} filtering via first-order multi-target moments.
\newblock {\em IEEE Trans. Aerosp. Electron. Syst.}, 39(4):1152--1178, 2003.

\bibitem{Mah03a}
R.~Mahler.
\newblock Objective functions for {Bayesian} control-theoretic sensor
  management, {I}: Multitarget first-moment approximation.
\newblock In {\em Proc. IEEE Aerospace Conference}, volume~4, pages 1905--1923,
  2003.

\bibitem{Mah04}
R.~Mahler.
\newblock Multitarget sensor management of dispersed mobile sensors.
\newblock In D.~Grundel, R.~Murphey, and P.~Pardalos, editors, {\em Theory and
  algorithms for cooperative systems}, chapter~12, pages 239--310. World
  Scientific Books, 2004.

\bibitem{MahlerCPHD}
R.~Mahler.
\newblock {P}{H}{D} filters of higher order in target number.
\newblock {\em IEEE Trans. Aerosp. Electron. Syst.}, 43(4):1523--1543, 2007.

\bibitem{Mah07}
R.~Mahler.
\newblock {\em Statistical Multisource-Multitarget Information Fusion}.
\newblock Artech House, 2007.

\bibitem{RAG04}
B.~Ristic, S.~Arulampalam, and N.~Gordon.
\newblock {\em Beyond the Kalman filter: Particle filters for tracking
  applications}.
\newblock Artech House, 2004.

\bibitem{RV10}
B.~Ristic and B-N. Vo.
\newblock Sensor control for multi-object state-space estimation using random
  finite sets.
\newblock {\em Automatica}, 46(11):1812--1818, 2010.

\bibitem{RVC11}
B.~Ristic, B-N. Vo, and D.~Clark.
\newblock A note on the reward function for {PHD} filters with sensor control.
\newblock {\em IEEE Trans. Aerosp. Electron. Syst.}, 47(2):1521--1529, 2011.

\bibitem{SVV08}
D.~Schumacher, B-T. Vo, and B.-N. Vo.
\newblock A consistent metric for performance evaluation of multi-object
  filters.
\newblock {\em IEEE Trans. Signal Process.}, 56(8):3447--3457, 2008.

\bibitem{SKV07}
S.~Singh, N.~Kantas, B.-N. Vo, A.~Doucet, and R.~Evans.
\newblock Simulation based optimal sensor scheduling with application to
  observer trajectory planning.
\newblock {\em Automatica}, 43(5):817--830, 2007.

\bibitem{SKM95}
D.~Stoyan, D.~Kendall, and J.~Mecke.
\newblock {\em Stochastic Geometry and its Applications}.
\newblock John Wiley \& Sons, 1995.

\bibitem{VoMaGMPHD05}
B.-N. Vo and W.-K. Ma.
\newblock The {G}aussian mixture probability hypothesis density filter.
\newblock {\em IEEE Trans. Signal Process.}, 54(11):4091--4104, 2006.

\bibitem{VoAES}
B.-N. Vo, S.~Singh, and A.~Doucet.
\newblock Sequential {Monte Carlo} methods for multi-target filtering with
  random finite sets.
\newblock {\em IEEE Trans. Aerosp. Electron. Syst.}, 41(4):1224--1245, 2005.

\bibitem{VoGaussianCPHD07}
B.-T. Vo, B.-N. Vo, and A.~Cantoni.
\newblock Analytic implementations of the cardinalized probability hypothesis
  density filter.
\newblock {\em IEEE Trans. Signal Process.}, 55(7):3553--3567, 2007.

\bibitem{VVC09}
B.-T. Vo, B.-N. Vo, and A.~Cantoni.
\newblock The cardinality balanced {Multi-Target Multi-Bernoulli} filter and
  its implementations.
\newblock {\em IEEE Trans. Signal Process.}, 57(2):409--423, 2009.

\end{thebibliography}

\end{document}